\documentclass[final,5p,twocolumn]{elsarticle}

\usepackage{amssymb}
\usepackage{amsmath}

\usepackage{overpic}

\biboptions{numbers,square,sort&compress}

\begin{document}

\begin{frontmatter}



\title{Open heavy flavor in Pb+Pb collisions at $\sqrt{s}$=2.76 TeV within a transport model}


\author[label1]{Jan Uphoff}
\author[label1]{Oliver Fochler}
\author[label2]{Zhe Xu}
\author[label1]{Carsten Greiner}

\address[label1]{Institut f\"ur Theoretische Physik, Johann Wolfgang 
Goethe-Universit\"at Frankfurt, Max-von-Laue-Str. 1, \\
D-60438 Frankfurt am Main, Germany}
\address[label2]{Department of Physics, Tsinghua University, Beijing 100084, China}

\begin{abstract}
The space-time evolution of open heavy flavor is studied in Pb+Pb collisions at $\sqrt{s}$=2.76 TeV using the partonic transport model \emph{Boltzmann Approach to MultiParton Scatterings} (BAMPS). An updated version of BAMPS is presented which allows interactions among all partons: gluons, light quarks and heavy quarks. Heavy quarks, in particular, interact with the rest of the medium via binary scatterings with a running coupling and a Debye screening which is matched by comparing to hard thermal loop calculations. The lack of radiative processes in the heavy flavor sector is accounted for by scaling the binary cross section with a phenomenological factor $K=3.5$, which describes well the elliptic flow $v_2$ and nuclear modification factor $R_{AA}$ at RHIC. Within this framework we calculate in a comprehensive study the $v_2$ and $R_{AA}$ of all interesting open heavy flavor particles at the LHC: electrons, muons, $D$ mesons, and non-prompt $J/\psi$ from $B$ mesons. We compare to experimental data, where it is already available, or make predictions. To do this accurately next-to-leading order initial heavy quark distributions are employed which agree well with proton-proton data of heavy flavor 
at  $\sqrt{s}$=7 TeV.
\end{abstract}

\begin{keyword}
heavy quarks \sep open heavy flavor \sep elliptic flow  \sep nuclear modification factor \sep LHC


\end{keyword}

\end{frontmatter}


\section{Introduction}

In ultra-relativistic heavy ion collisions at the BNL RHIC and CERN LHC the energy deposited in the collision zone is large enough to produce a medium that consists of deconfined quarks and gluons \cite{Adams:2005dq,Adcox:2004mh,Muller:2012zq}. This  state of matter, the quark gluon plasma (QGP), has remarkable properties such as collective behavior, a small viscosity to entropy ratio, and a large density that leads to quenching of jets.

Open heavy flavor particles such as $D$ and $B$ mesons, which consists of one heavy quark and one light quark, are an exciting probe to study the QGP. Their heavy constituents, namely charm and bottom quarks, are created at a very early stage of the heavy ion collision due to their large mass \cite{Uphoff:2010sh}. Consequently, they travel for a long time through the QGP, collide with other medium particles, lose energy, and participate in the collective behavior.

Measurements of heavy flavor electrons from the decay of $D$ and $B$ mesons at RHIC  \cite{Abelev:2006db,Adare:2006nq,Adare:2010de} indicate that the elliptic flow $v_2$ and nuclear modification factor $R_{AA}$ of open heavy flavor is on the same order as for light particles. This is in contrast to the expectations drawn from the dead cone effect \cite{Dokshitzer:2001zm,Abir:2011jb}, that gluon radiation off heavy quarks is suppressed at small angles compared to light quarks. The reason for the rather large elliptic flow and suppression of heavy flavor is currently under investigation \cite{Armesto:2005mz,vanHees:2005wb,Moore:2004tg,Mustafa:2004dr,Wicks:2005gt,Zhang:2005ni,Adil:2006ra,Molnar:2006ci,Peigne:2008nd,Gossiaux:2008jv,Alberico:2011zy,Uphoff:2011ad,Abir:2012pu}.

At the LHC, for the first time, it is possible to distinguish between charm and bottom quarks. In addition to looking at heavy flavor electrons or muons from both $D$ and $B$ mesons, ALICE can reconstruct $D$ mesons directly~\cite{ALICE:2012ab}. In addition, CMS presented data for non-prompt $J/\psi$ \cite{Chatrchyan:2012np} which stem from the decay of $B$ mesons. 

In the present paper we present our calculations of the elliptic flow and nuclear modification factor of $D$ mesons and non-prompt $J/\psi$, obtained with the transport model \emph{Boltzmann Approach to MultiParton Scatterings} (BAMPS). In addition, we show results of heavy flavor muons at forward rapidity as well as electrons at mid-rapidity and compare them to experimental data.

\section{Open heavy flavor in BAMPS}
In this section we briefly review the features of our model, the parton cascade \emph{Boltzmann Approach to MultiParton Scatterings} (BAMPS). More details concerning the model itself and the implementation of heavy flavor can be found in  Ref.~\cite{Xu:2004mz,Xu:2007aa} and \cite{Uphoff:2011ad,Uphoff:2010sh}, respectively.

BAMPS is a 3+1 dimensional partonic transport model that solves the Boltzmann equation
\begin{equation}
\label{boltzmann}
\left ( \frac{\partial}{\partial t} + \frac{{\mathbf p}_i}{E_i}
\frac{\partial}{\partial {\mathbf r}} \right )\,
f_i({\mathbf r}, {\mathbf p}_i, t) = {\cal C}_i^{2\rightarrow 2} + {\cal C}_i^{2\leftrightarrow 3}+ \ldots  \ ,
\end{equation}
for on-shell partons. Implemented processes on the light parton sector are all $2 \rightarrow 2$ and $2 \leftrightarrow 3$ processes. In contrast to previous publications \cite{Uphoff:2011ad,Uphoff:2011aa,Uphoff:2010bv,Fochler:2011en} where we only took gluons ($g$) and heavy quarks (Q) into account, in the present calculation light quarks ($q$) are explicitly included.
All cross sections are calculated in leading order pQCD. Light partons interact among each other via binary collisions and radiative $2 \leftrightarrow 3$ processes, which are taken in the Gunion-Bertsch limit \cite{Gunion:1981qs}.

For heavy quarks, currently only the elastic collisions
\begin{align}
\label{heavy_quark_processes}
	      g g \leftrightarrow Q  \bar{Q} \nonumber\\
        q  \bar{q} \leftrightarrow Q  \bar{Q} \nonumber\\
        g Q \rightarrow g Q  \nonumber\\
        q Q \rightarrow q Q 
\end{align}
are implemented. The inclusion of radiative processes is underway and planned for the near future.

In this paper we focus on the heavy flavor sector. For BAMPS results of light partons we refer to Refs.~\cite{Xu:2004mz,Xu:2007aa,Xu:2007jv,Xu:2007ns,Xu:2008av,Fochler:2008ts,Xu:2010cq,Fochler:2010wn,Fochler:2011en}.

The cross sections for the processes from \eqref{heavy_quark_processes} are calculated in leading order pQCD for a finite heavy quark mass \cite{Combridge:1978kx}. Since the matrix elements  of the $t$ channel of those elastic heavy quark scattering with a light parton are divergent, they are screened with a screening mass, which is determined from comparison to hard thermal loop (HTL) calculations. By comparing the energy loss of a heavy quark in a static medium calculated within HTL and the same quantity calculated from the leading order pQCD cross section with a screening mass $\mu^2 = \kappa m_D^2$, one can obtain the  prefactor $\kappa$ analytically to be \cite{Gossiaux:2008jv,Peshier:2008bg,Uphoff:2011ad}
\begin{align}
	\kappa = \frac{1}{2 {\rm e}} \approx 0.184 \approx 0.2 \ .
\end{align}
The Debye mass $m_D^2$ is calculated in BAMPS from the non-equilibrium distribution functions $f$ of gluons and light quarks via
\cite{Xu:2004mz}  
\begin{equation}
\label{md2}
m_D^2 = \pi \alpha_s \nu_g \int \frac{{\rm d}^3p}{(2\pi)^3} \frac{1}{p} 
( N_c f_g + n_f f_q) \ ,
\end{equation}
where $N_c=3$ denotes the number of colors, $\nu_g = 16$ is the gluon degeneracy, and $n_f = 3$ the number of active light flavors. As a note, in equilibrium and with Boltzmann statistics the Debye mass is given by $m_{D,{\rm eq}}^2 = \frac{8 \alpha_s}{\pi} (N_c+n_f) \, T^2$. The Debye mass prefactor $\kappa$ was determined for an arbitrary number of light quark degrees of freedom $n_f$ and is thus easily applied to the $t$ channel of heavy quark interactions with light quarks.

Furthermore, instead of just assuming a constant value for the coupling $\alpha_s$ we employ the running coupling for all heavy flavor processes \cite{Dokshitzer:1995qm,Gossiaux:2008jv,Peshier:2008bg,Uphoff:2011ad},
\begin{align}
\label{alpha_s_continued}
 \alpha_s(Q^2)= \frac{4\pi}{\beta_0} \begin{cases}
  L_-^{-1}  & Q^2 < 0\\
  \frac12 - \pi^{-1} {\rm arctan}( L_+/\pi ) &  Q^2 > 0
\end{cases}
\end{align}
with $\beta_0 = 11-\frac23\, n_f$  and
$L_\pm = \ln(\pm Q^2/\Lambda^2)$ with $\Lambda=200 \, {\rm MeV}$.
For consistency a running coupling is also used in calculating the Debye mass from Eq.~\eqref{md2}.

After the energy density in the surrounding of a heavy quark in BAMPS has dropped below $0.6 \, {\rm GeV/fm^3}$ it is fragmented to a $D$ or $B$ meson. This is done according to the Peterson fragmentation function \cite{Peterson:1982ak}
\begin{align}
D_{H/Q} (z) = \frac{N}{z \left( 1 - \frac{1}{z} - \frac{\epsilon_Q}{ 1 - z } \right)^2} \ .
\end{align}
$N$ is a normalization constant, $z = |\vec{p}_H|/|\vec{p}_Q|$ the ratio of the meson and quark momenta, and $\epsilon_Q = 0.05$~($0.005$) for charm (bottom) quarks.
The $D$ mesons can then directly be compared to the experimental data. To yield non-prompt $J/\psi$ we carry out the decay of $B$ mesons with \textsc{pythia} \cite{Sjostrand:2006za,Sjostrand:2007gs} by switching on the relevant decay channels. \textsc{pythia} is also used to perform the decay of $D$ and $B$ mesons to electrons and muons which can then be compared to experimental data.

Especially for the LHC, where charm and bottom can be separated, it is important to have the correct reference for the initial heavy quark distribution. To generate the initial heavy quark spectrum for the BAMPS simulation of the heavy ion collision we employ the next-to-leading order (NLO) event generator \textsc{mc@nlo} \cite{Frixione:2002ik,Frixione:2003ei}. The factorization  and renormalization scales, $\mu_F$ and $\mu_R$, respectively, are in principle arbitrary when considering all orders of the cross section. However, for the leading order cross section uncertainties due to neglecting higher order terms can be reduced if the two scales are of the order of the relevant scale $\sqrt{p_T^2 + M^2}$, $p_T^2$ being the transverse momentum and $M$ the mass of the produced heavy quarks. The exact value of the scale is fixed by giving a good agreement with the experimental data which results in $\mu_F=\mu_R = 1\, \sqrt{p_T^2 + M_c^2}$ for charm ($M_c=1.3 \, {\rm GeV}$) and $\mu_F=\mu_R = 0.4\, \sqrt{p_T^2 + M_b^2}$ for bottom quarks ($M_b=4.6 \, {\rm GeV}$).

\begin{figure}
\includegraphics[width=1.0\linewidth]{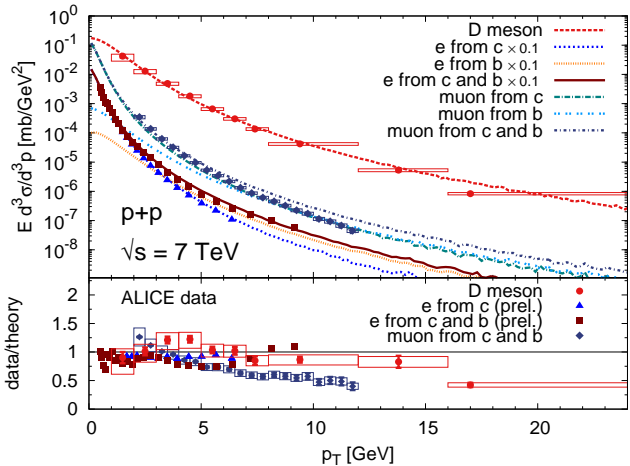}
\caption{Differential invariant cross section of $D$ mesons with $|y|<0.5$ and heavy flavor electrons with $|y|<0.8$ at mid-rapidity and muons at forward rapidity $2.5<y<4$ as a function of transverse momentum for proton-proton collisions with $\sqrt{s}=7 \, {\rm TeV}$ simulated with \textsc{mc@nlo}. For comparison experimental data \cite{ALICE:2011aa, Pachmayer:2011qu, Abelev:2012pi} with the same kinematic cuts is also shown. In the upper plot the electron curves and the corresponding data points have been scaled with the factor~0.1 to distinguish them from the muon curves.
Since the data of electrons is preliminary, we  do not have access to the errors and plot those data points without any errors as obtained from Ref.~\cite{Pachmayer:2011qu}.
}
\label{fig:pp_mcatnlo}
\end{figure}

In Fig.~\ref{fig:pp_mcatnlo} the invariant differential cross sections of $D$ mesons, heavy flavor electrons and muons are compared to experimental data from ALICE at $\sqrt{s}=7 \, {\rm TeV}$.
The $D$ mesons and heavy flavor electrons at mid-rapidity are well described by \textsc{mc@nlo}. At forward rapidity, however, the slope of the muons at larger $p_T$ is slightly different. Such a disagreement has also been observed by CMS in a more detailed study of inclusive bottom jets in Ref.~\cite{Chatrchyan:2012dk} by comparing \textsc{mc@nlo} to data for larger $p_T$ and various rapidities. Nevertheless, we checked that the muon $R_{AA}$ is not very sensitive to the exact slope in this $p_T$ range. 

To obtain the initial heavy quark distribution as an input for BAMPS we run \textsc{mc@nlo} with the same parameters for a center of mass energy of $\sqrt{s}=2.76 \, {\rm TeV}$ which was used for the recent heavy ion runs.

\section{Results}
\label{sec:results}

Important 
observables for open heavy flavor are the nuclear modification factor $R_{AA}$ and the elliptic flow $v_2$. The nuclear modification factor is defined as the heavy flavor yield in heavy ion collisions divided by the yield from p+p collisions scaled with the number of binary collisions,
\begin{align}
  R_{AA}=\frac{{\rm d}^{2}N_{AA}/{\rm d}p_{T}{\rm d}y}{N_{\rm bin} \, {\rm d}^{2}N_{pp}/{\rm d}p_{T}{\rm d}y} \ .
\end{align} 
The elliptic flow denotes the second harmonic of the Fourier decomposition of the azimuthal particle spectrum and is given by
\begin{align}
\label{elliptic_flow}
  v_2=\left\langle  \frac{p_x^2 -p_y^2}{p_T^2}\right\rangle \ .
\end{align} 
For this definition the momenta $p_x$ and $p_y$ are taken with respect to the reaction plane.

\begin{figure}[t]
\includegraphics[width=1.0\linewidth]{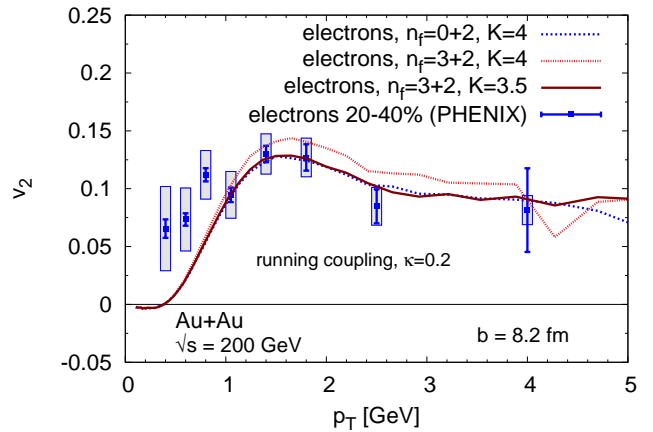}
\caption{Elliptic flow $v_2$ of heavy flavor electrons at RHIC with an impact parameter $b=8.2 \, {\rm fm}$ together with data \cite{Adare:2010de}. For heavy quarks only binary collisions are switched on, which are multiplied with a factor $K$.
}
\label{fig:rhic_v2}
\end{figure}

\begin{figure}[t]
\includegraphics[width=1.0\linewidth]{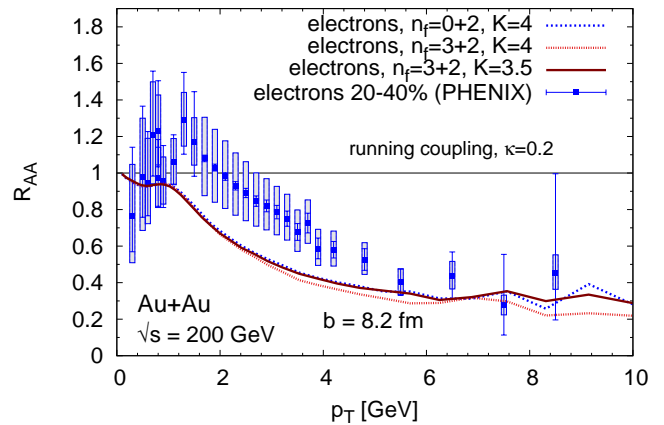}
\caption{Nuclear modification factor $R_{AA}$ of heavy flavor electrons at RHIC for the same configuration as in Fig.~\ref{fig:rhic_v2}.
}
\label{fig:rhic_raa}
\end{figure}

In previous publications \cite{Uphoff:2011ad,Uphoff:2011aa,Uphoff:2010bv} we showed that elastic heavy quark scatterings alone cannot describe the experimental data at RHIC, although they play a significant role. To be compatible with the heavy flavor electron data for $R_{AA}$ and $v_2$ at RHIC the elastic cross section had to be multiplied with a factor $K=4$.  Those calculations have been done without light quarks, $n_f=0+2$ (number of light quarks $+$ heavy quarks). Figures~\ref{fig:rhic_v2} and \ref{fig:rhic_raa} compare these calculations to the updated version of BAMPS which also includes light quarks ($n_f=3+2$).
Comparing the $K=4$ curves for $n_f=0+2$ and $n_f=3+2$ shows that both the suppression and elliptic flow are slightly higher for the latter. At first sight this seems counterintuitive since we use the same initial conditions from \textsc{pythia} for both cases and, to get the same energy density, convert initial light quarks to gluons for $n_f=0+2$, which are associated with a larger Casimir factor. However, the situation is more complex. The running coupling also depends on the number of flavors and is larger for $n_f=3+2$. Furthermore, the chemistry of a purely gluonic plasma behaves slightly differently as a quark gluon plasma. This leads to a larger number of scattering centers of the medium for $n_f=3+2$. All influences together result in a slight increase of the suppression and elliptic flow for $n_f=3+2$ compared to $n_f=0+2$.

As can be seen in Figs.~\ref{fig:rhic_v2} and \ref{fig:rhic_raa} the best agreement with data for $n_f=3+2$ is found with $K=3.5$ in contrast to $n_f=0+2$, where $K=4$ yielded the best results.
We assume that this factor is necessary due to the lack of radiative processes and quantum statistics in our calculations. It is planned to extend BAMPS in a forthcoming study to include also radiative processes. This will show if those processes can indeed account for such a phenomenological scaling of the binary cross section. As a note, the value of 3.5 is close to the needed $K$ factor of Ref.~\cite{Meistrenko:2012ju}, which, in an independent framework, included similar cross sections for $n_f=3+2$ using ideal hydro as well as temperature and flow information from BAMPS for the medium evolution.

For small $p_T$ the employed hadronization scheme, namely Peterson fragmentation, is not valid and coalescence might be the dominant process. It is expected that coalescence increases the elliptic flow at small transverse momenta, since light quarks would also contribute to the flow of the heavy mesons. This could be an explanation why BAMPS underestimates the flow in Fig.~\ref{fig:rhic_v2} for very small $p_T$. Neglecting cold nuclear matter effects such as the Cronin effect or shadowing and also coalescence effects  could be the reason for the deviation of the $R_{AA}$ for small $p_T$ in Fig.~\ref{fig:rhic_raa}.

In conclusion, the effective description of the RHIC data with $K=3.5$ agrees simultaneously with the data for both $R_{AA}$ and $v_2$ for intermediate and large $p_T$. In this paper we will employ this prescription at LHC energy of $\sqrt{s}=2.76 \, {\rm TeV}$ and compare to experimental data for $D$ meson $R_{AA}$ and $v_2$, non-prompt $J/\psi$, electron and muon $R_{AA}$. Furthermore, we make predictions for non-prompt $J/\psi$, electron, and muon $v_2$.
As a note, for all calculations the same kinematic acceptance cuts are set at which experimental data is measured (see labels in the plots for  the values).
The impact parameters used in BAMPS are matched with a Glauber calculation to the mean number of participants $\left\langle N_{\rm part}\right\rangle$ given for each centrality class \cite{Aamodt:2010cz}.

\begin{figure}
\includegraphics[width=1.0\linewidth]{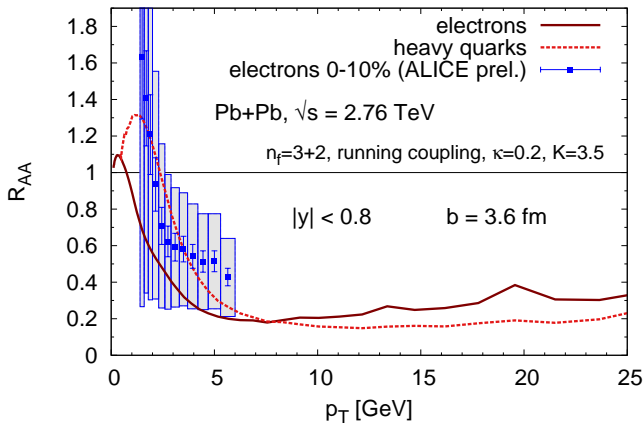}
\caption{Nuclear modification factor $R_{AA}$ of heavy flavor electrons at Pb+Pb collisions at LHC with an impact parameter $b= 3.6 \, {\rm fm}$ together with data \cite{Pachmayer:2011wq}. For heavy quarks only binary collisions are switched on, which are multiplied with $K=3.5$.}
\label{fig:raa_lhc_electron}
\end{figure}

In Ref.~\cite{Uphoff:2011ad} we presented predictions for the electron $R_{AA}$ and $v_2$ at LHC for $n_f=0+2$. In Fig.~\ref{fig:raa_lhc_electron} an update of the heavy flavor electron $R_{AA}$ to $n_f=3+2$ is shown, employing again, as at RHIC, only collisional energy loss with a running coupling, improved Debye screening, and $K=3.5$. 
Our calculation is consistent with the preliminary experimental data, although it is on the lower edge of the uncertainty band. However, due to the rather large error bars one cannot judge yet whether the effective description of the RHIC data also applies at the LHC. 

One huge advantage of transport models is the direct access to all particles of the QGP during the whole time evolution. Therefore, we plot in addition to the heavy flavor electron curve the $R_{AA}$ of charm and bottom quarks. As can be seen in Fig.~\ref{fig:raa_lhc_electron} the curve of the heavy flavor electrons is shifted to smaller $p_T$ compared to the parental heavy quarks due to fragmentation and decay processes. Predictions for electron elliptic flow will be presented along with muons and non-prompt $J/\psi$ at the end of the section since there is no experimental data available yet.

\begin{figure}
\includegraphics[width=1.0\linewidth]{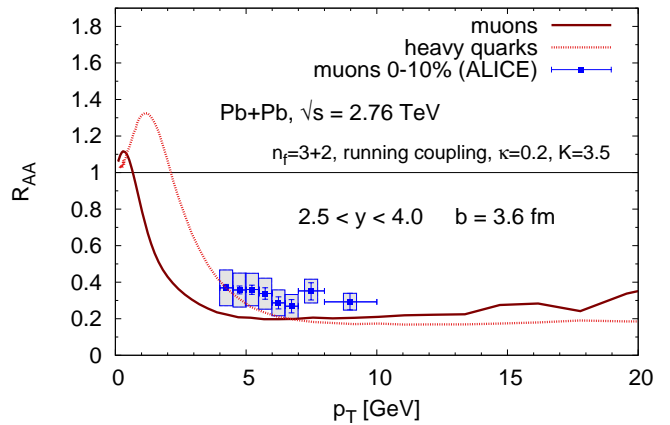}
\caption{$R_{AA}$ of muons at forward rapidity at LHC with data \cite{Abelev:2012qh}.}
\label{fig:raa_lhc_muon}
\end{figure}

ALICE can measure muons in forward rapidity stemming from heavy flavor decays. Figure~\ref{fig:raa_lhc_muon} shows the BAMPS results in the same rapidity range as the experimental data.
Comparing to Fig.~\ref{fig:raa_lhc_electron}, it is obvious that the suppression of muons at forward rapidity is as strong as that of electrons at mid-rapidity and both nuclear modification factors assume very similar values. In contrast to the electron data, the muon data has only small errors and a deviation is visible between the data and our curve, which is calculated for the same parameters that describe the RHIC heavy flavor electron data. In addition to the muon curve, we show the $R_{AA}$ on the heavy quark level directly which is again shifted to larger $p_T$ as one would expect.

By considering muons or electrons, the contributions from charm and bottom cannot be distinguished. However, at the LHC for the first time one has access to charm and bottom separately via $D$ mesons and non-prompt $J/\psi$ from $B$ mesons.

\begin{figure}[ht]
\includegraphics[width=1.0\linewidth]{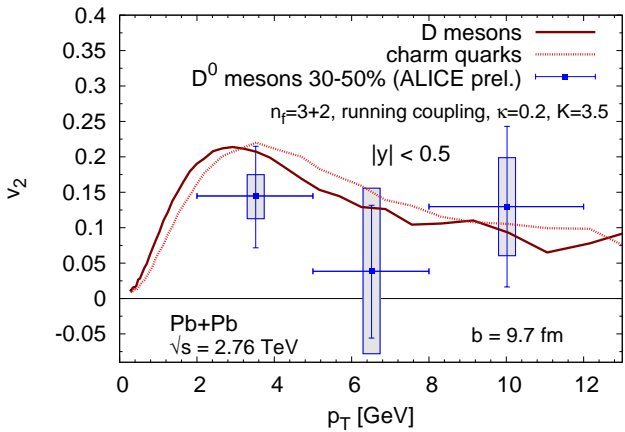}
\caption{Elliptic flow $v_2$ of $D$ mesons at Pb+Pb collisions at LHC with an impact parameter $b = 9.7 \, {\rm fm}$ together with data \cite{Bianchin:2011fa}.
}
\label{fig:v2_d_meson_lhc}
\end{figure}

Figure~\ref{fig:v2_d_meson_lhc} depicts the BAMPS calculations for $D$ meson $v_2$.
The error bars of the preliminary data are too large to draw any definite conclusion, but our results are in good agreement within the errors. In addition to the $D$ meson curve we plot the charm quark $v_2$. The difference between both curves is considerably smaller than it was the case for heavy flavor electrons or muons, which renders $D$ mesons a very good indicator for actual charm quark observables.

\begin{figure}
\includegraphics[width=1.0\linewidth]{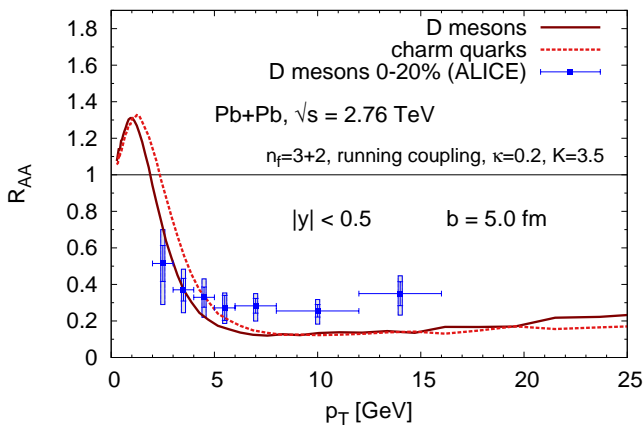}
\caption{Nuclear modification factor $R_{AA}$ of $D$ mesons at LHC with data \cite{ALICE:2012ab}.}
\label{fig:raa_d_meson_lhc}
\end{figure}

The $D$ meson $R_{AA}$ from BAMPS is compared to data in Fig.~\ref{fig:raa_d_meson_lhc}. Although the order of magnitude of the suppression is comparable, the experimental data tends to be slightly underestimated by our calculation. 
This is in accordance to the muon  $R_{AA}$ at forward rapidity in Fig.~\ref{fig:raa_lhc_muon}, which is also below the data, and the electron $R_{AA}$ in Fig.~\ref{fig:raa_lhc_electron}, which is at the lower edge of the error bars.
This could be a first hint that new effects compared to RHIC play a role at the LHC. An indication in this direction is also the fact that $D$ meson suppression seems to be slightly smaller than that of charged hadrons \cite{ALICE:2012ab}.

Possible explanations for the discrepancy in our $R_{AA}$ calculations and the heavy flavor data could be cold nuclear matter effects, the normalization error of the data which is not shown in the plot or that we represent the rather large centrality classes by only one impact parameter. 
Furthermore, a reason could be that the approximation of modeling the radiative energy loss by scaling the binary cross section with a constant factor is not satisfied. Although we do not expect that such a $K$ factor is temperature dependent for a thermalized system, non-thermal effects in the medium evolution could trigger different $K$ factors for different collision energies at RHIC and LHC. However, this cannot be assessed without actually doing the calculation with higher order processes where the $K$ factor is obsolete.  We will investigate this in more detail in a forthcoming study.

\begin{figure}
\includegraphics[width=1.0\linewidth]{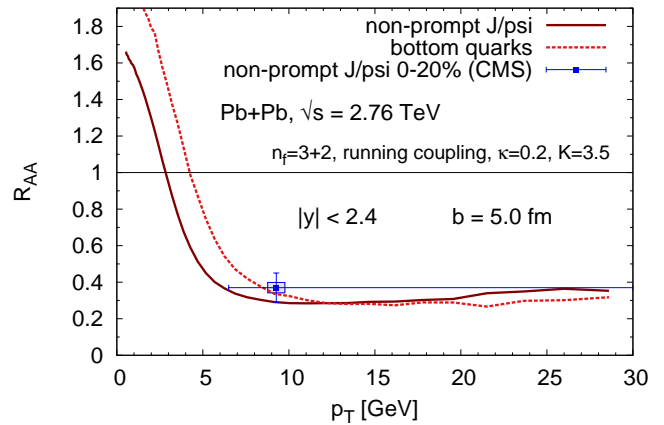}
\caption{$R_{AA}$ of non-prompt $J/\psi$ at LHC with data \cite{Chatrchyan:2012np}.}
\label{fig:raa_lhc_npjpsi}
\end{figure}

A complimentary measurement has been performed by the CMS collaboration \cite{Chatrchyan:2012np} which measured the suppression of non-prompt $J/\psi$ from the decay of $B$ quarks. Although only one data point could be extracted, the suppression of non-prompt $J/\psi$ is clearly visible in Fig.~\ref{fig:raa_lhc_npjpsi} and the magnitude is in good agreement with our calculation.
Analogously to the other $R_{AA}$ comparisons at LHC, our curve is slightly smaller than the experimental value, although still within the errors. Again, the suppression of bottom quarks themselves is very similar to that of non-prompt $J/\psi$.

\begin{figure}
\includegraphics[width=1.0\linewidth]{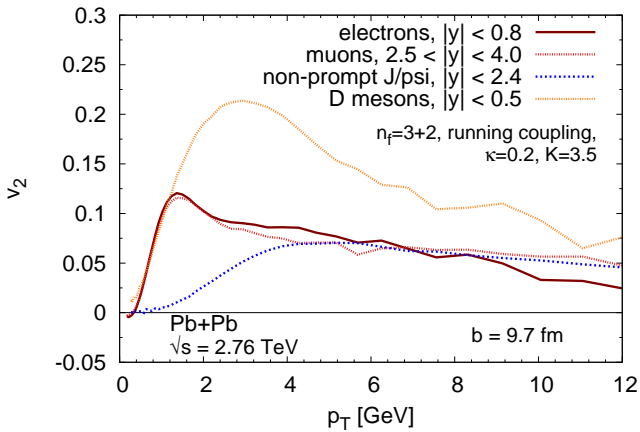}
\caption{Predictions for the elliptic flow $v_2$ of muons, electrons, non-prompt $J/\psi$, and $D$ mesons.
}
\label{fig:lhc_v2_muon_e_npj}
\end{figure}

To conclude the $v_2$ and $R_{AA}$ comparisons we show in Fig.~\ref{fig:lhc_v2_muon_e_npj} BAMPS predictions of the elliptic flow of muons, electrons, and non-prompt $J/\psi$ calculated with the same parameters used for the previous figures, which describe the RHIC data.
For better comparison the curve of $D$ mesons from Fig.~\ref{fig:v2_d_meson_lhc} is also depicted. 

The flow of non-prompt $J/\psi$ is considerably smaller than the $D$ meson flow due to the mass difference of charm and bottom quarks. Accordingly, the influence of bottom quarks to the flow of heavy flavor electrons at intermediate and large $p_T$ is also the reason why the electron flow does not increase to the value of the $D$ meson flow. Muons at forward rapidity adopt the same elliptic flow as electrons at mid-rapidity. This is in accordance with the same $R_{AA}$ of muons and electrons (cf. Figs.~\ref{fig:raa_lhc_muon} and \ref{fig:raa_lhc_electron}). 
Since BAMPS is a 3+1 dimensional transport model, boost invariance of the system in rapidity is not assumed, but -- in first approximation -- comes out naturally for not too large rapidity gaps, which is reflected in the same $v_2$ and $R_{AA}$ of electrons and muons at mid- and forward rapidity, respectively.

\begin{figure*}[ht]
\begin{minipage}[t]{0.49\textwidth}
\centering
\includegraphics[width=1.0\textwidth]{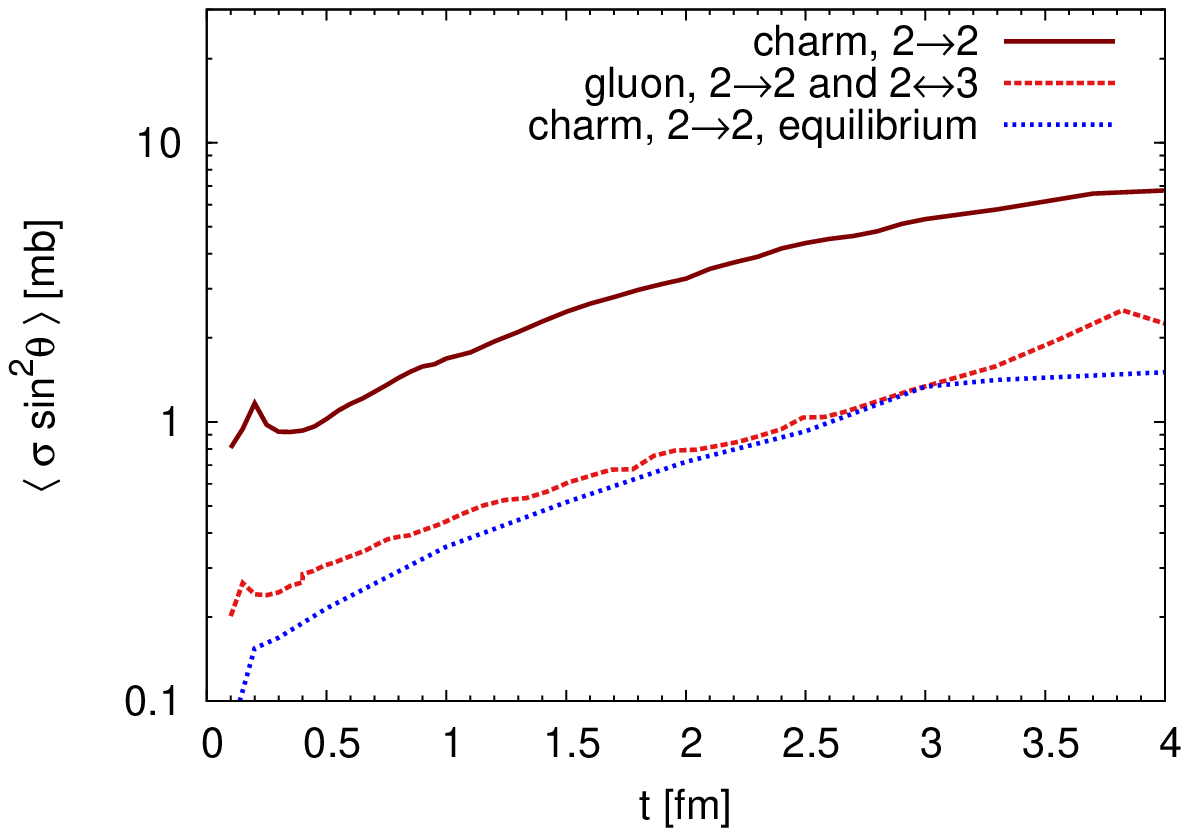}
\end{minipage}
\hfill
\begin{minipage}[t]{0.49\textwidth}
\centering
\includegraphics[width=1.0\textwidth]{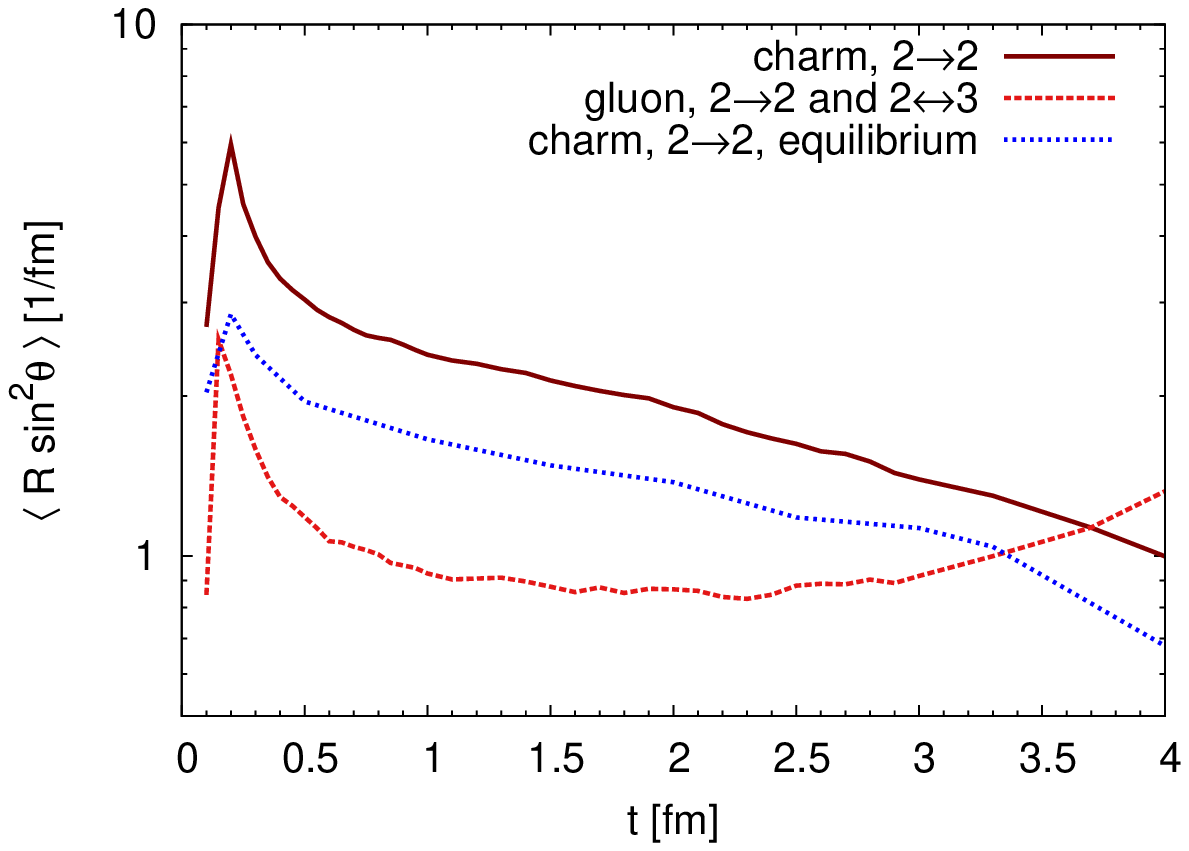}
\end{minipage}
\caption{Mean transport cross section $\langle \sigma \sin^2 \theta  \rangle$ (left) and transport rate $\langle R \sin^2 \theta  \rangle$ (right) of charm quarks and gluons as a function of time in the central region of a heavy ion collision at LHC ($b=9.7\, {\rm fm}$, $\sqrt{s}=2.76 \, {\rm TeV}$, $n_f= 3+2$, $K=3.5$ for charm quarks). $\theta$ is the angle between the momenta of the considered charm quark (gluon) before and after the collision in the lab frame. The cross section is averaged over all particles in a tube with a radius of $1.5 \, {\rm fm}$ and space-time rapidity $\eta \in [-0.5;0.5]$. In addition, the charm transport cross section and rate  in a chemically equilibrated medium is shown. That is, for each time we extract the medium energy density and mean charm energy from BAMPS in the central region and compute the transport cross section and rate of a charm quark with that energy in a static and equilibrated medium.}
\label{fig:transport_cs}
\end{figure*}

Finally, the question arises, what we can learn from the comparison to experimental data. Due to the transport character of BAMPS we have access to all the collision properties during the whole time evolution. 
Figure~\ref{fig:transport_cs} sheds some light on why the experimental data can be fairly well described with the $K$ factor.

For the elliptic flow isotropization is important and, hence, the transport cross section and transport rate are the relevant quantities, since they weight the cross section and rate, respectively, with the angle of the diffracted particle. In the left plot of Fig.~\ref{fig:transport_cs} the time evolution of the mean transport cross section is shown in the central region of a heavy ion collision at LHC. The value for charm quarks, including only $2 \rightarrow 2$ processes, is about 5 times larger than that for gluons, which interact also via $2 \leftrightarrow 3$ processes. As a note, the  $2 \rightarrow 2$ and $2 \leftrightarrow 3$ light parton processes can describe the elliptic flow data of light particles \cite{Xu:2007jv,Xu:2008av,Xu:2010cq}. 
Due to the large charm transport cross section also a sizeable elliptic flow for charm quarks builds up.

For comparison we show also the charm transport cross section under the assumption that the medium is thermally and chemically equilibrated. These values are smaller compared to the full heavy ion collision, where the gluon fugacity is below unity, which leads to a smaller Debye mass and, therefore, larger cross section.
However, the sensitivity on the fugacity is reduced when considering the transport rates since the density also enters. The right hand side of Fig.~\ref{fig:transport_cs} shows that for the charm transport rate the equilibrium curve is only a factor of about 1.4 smaller than the values extracted from the full heavy ion collision.

The reason for the large charm cross section is the effective treatment of the Debye screening on the heavy flavor sector and the additional multiplication of the factor $K=3.5$. Since the binary pQCD cross section is dominated by small angles, the total cross section can be one order larger than the transport cross section, which is too large for partonic interactions. Therefore, we will investigate in a further study radiative processes, which can  lead to an effective energy loss and isotropization (cf. also Ref.~\cite{Abir:2012pu}) without the need of large cross sections, as it has been demonstrated with gluons and light quarks in BAMPS \cite{Xu:2004mz}.

\section{Conclusions}
Results on the elliptic flow and the nuclear modification factor of open heavy flavor at the LHC are calculated within the partonic transport model BAMPS with $n_f = 3+2$ flavors. To get the correct initial heavy quark distribution we employ \textsc{mc@nlo} which is in good agreement with p+p heavy flavor data at $\sqrt{s}=7 \, {\rm TeV}$. The binary cross sections of heavy quarks with light medium particles are obtained from pQCD with a running coupling and an improved Debye screening motivated from hard thermal loop calculations. By scaling the binary cross section by a factor of $K=3.5$ to account for radiative contributions and quantum statistics we find a good agreement with the heavy flavor electron data at RHIC. Calculations with the same parameters are carried out at the LHC for heavy flavor electrons, muons, $D$ mesons and non-prompt $J/\psi$. 
We find that our $R_{AA}$ calculations slightly underestimate the experimental data for all heavy flavor observables. In a future study we want to investigate this in more detail by explicitly including radiative processes for heavy quarks \cite{Abir:2011jb} and study whether a phenomenological $K$ factor was satisfied. The $D$ meson $v_2$ agrees well with the experimental data. Furthermore, we make predictions within the same framework for the elliptic flow of heavy flavor electrons, muons, and non-prompt $J/\psi$ at LHC energy of $\sqrt{s}=2.76 \, {\rm TeV}$.

\section*{Acknowledgements}
The BAMPS simulations were performed at the Center for Scientific Computing of the Goethe University Frankfurt. This work was supported by the Helmholtz International Center for FAIR within the framework of the LOEWE program launched by the State of Hesse.





\bibliography{hq}







\end{document}